\documentclass[lettersize,journal]{IEEEtran}
\usepackage{amsmath,amsfonts}
\usepackage{algorithmic}
\usepackage{algorithm}
\usepackage{array}
\usepackage{textcomp}
\usepackage{stfloats}
\usepackage{xurl}
\usepackage[colorlinks,urlcolor=blue,linkcolor=blue,citecolor=blue]{hyperref}
\usepackage{verbatim}
\usepackage{graphicx}
\usepackage{cite}
\usepackage{xurl}
\usepackage{subcaption}
\newcommand{\vect}[1]{\mathbf{#1}}

\begin{document}

\title{CoviChain: A Blockchain Based COVID-19 Vaccination Passport}

\author{%
Philip Bradish\textsuperscript{1}, Sarang Chaudhari\textsuperscript{2}, 
Michael Clear\textsuperscript{1} and Hitesh Tewari\textsuperscript{1}\\[1ex]
1 Trinity College Dublin \\
2 Indian Institute of Technology Delhi
}



\maketitle

\begin{abstract}
Vaccination passports are being issued by governments around the world in order to open up their travel and hospitality sectors. Civil liberty campaigners on the other hand argue that such mandatory instruments encroach upon our fundamental right to anonymity, freedom of movement, and are a backdoor to issuing “identity documents” to citizens by their governments. In this paper we present a privacy-preserving framework that uses two-factor authentication to create a unique identifier that can be used to locate a person’s vaccination record on a blockchain, but does not store any personal information about them. Our main contribution is the employment of a locality sensitive hashing algorithm over an iris extraction technique, that can be used to authenticate users and anonymously locate vaccination records on the blockchain, without leaking any personally identifiable information to the blockchain. Our proposed system allows for the safe reopening of society, while maintaining the privacy of citizens.
\end{abstract}

\begin{IEEEkeywords}
Blockchain, Vaccination Passport, Locality Sensitive Hashing, Biometrics, Privacy
\end{IEEEkeywords}

\section {Introduction}
The COVID-19 pandemic and the associated lockdowns have brought about unprecedented challenges to individuals, organizations and countries around the world. As we cautiously emerge from this pandemic on foot of a global vaccination drive, we face a follow-on challenge in the form of the contentious issue of ``vaccination passports". Vaccination passports (aka vaccination certificates) at present, are being issued by governments in various jurisdictions around the world in order to open up their travel and hospitality sectors. For some, vaccination passports represent a viable option for  getting back to some form of normality, but for many others, such instruments immediately bring back dark memories of ``Big Brother".

Civil liberty campaigners and many others argue that such mandatory instruments encroach upon our fundamental right to anonymity, freedom of movement, and are a backdoor to issuing ``identity documents" to citizens by their governments \cite{ACLU} \cite{Liberties}. They further argue that such documents will lead to discrimination and have serious implications for human rights, equality, data privacy etc. They fear that these documents may be issued as a temporary measure to protect the public during the pandemic, but might remain in place and become a permanent tool that helps various agencies to identify individuals.

It is difficult to argue against such concerns given the manner in which vaccination passports are being issued and made use of at the present time. These instruments are being issued in paper or digital form and contain huge amounts of personally identifiable information (PII), such as the owner's name, date of birth, gender, social security number etc. They reveal far more information about an individual than just their vaccination status. Also, they are prone to being misused by criminals, who may create forgeries of the certificates and sell them on to others. In extreme cases, the identity of users may be stolen and sold off to criminal gangs by hackers.

We believe that a more balanced approach is required, which places control in the hands of the owner of the certificate. It lets them decide whether they wish to allow someone to access their vaccination record, but at the same time does not reveal any PII about them to third parties. In this paper we present the design of a unique blockchain based COVID-19 vaccination passport framework (CoviChain) that fulfills the above requirements, and is versatile enough to be rolled out on a global scale.

The rest of the manuscript is organized as follows. We start by describing the overall CoviChain system. We then elaborate on each of the technical aspects of the system, such as the core algorithm, biometric authentication technique, the locality sensitive hashing mechanism we employ, and the blockchain operational details. Finally, we present the results and analysis of our simulations, and follow up with a number of security considerations.

\section{System Overview}
In order to develop a system that protects the PII of individuals, but at the same time stores their vaccination records anonymously and gives them total control over to whom and how the information can be disclosed, we make use of a two-factor authentication scheme. In particular, we make use of some characteristics that a user \textit{possesses} and some information that the user \textit{knows}.

For the \textit{possesses} part we make use of biometric information. Biometrics can be defined as a set of biological or physical measurements that make it possible to determine a person's identity by analysing physical traits such as voice, iris, face, fingerprints, etc. Over the past decade the availability of cheaper sensors has allowed for biometrics to be embedded in smartphones, and has evolved from simple fingerprint matching to more complex facial recognition systems \cite{Rui}. Biometrics are also increasingly being used by the banking and financial sectors to secure payment transactions \cite{Statista}. In particular, we make use of an iris scan in the CoviChain system.

For the \textit{knows} part we make use of personal details such as date of birth, gender, country of birth etc. We prefer to use personal details that can be entered into the system programmatically (e.g. a drop down menu), as opposed to manually entering them, as the latter is prone to error during the input process. We also make use of an immutable distributed ledger to securely store the vaccination records of each user registered with the system.

A person that wishes to obtain a vaccination certificate must present themselves to an authorized CoviChain center, where their iris is scanned and a feature vector is extracted - see section on \ref{biometric}. We make use of a locality sensitive hashing (LSH) function to obtain a hash of the feature vector - see section on \ref{lsh} \cite{Charikar}. The hash of the iris scan is then concatenated with the user's \textit{date of birth} and \textit{gender}, and the resultant quantity is then hashed again using the SHA-256 \cite{SHA} algorithm, to produce a unique identifier ($ID$) for the user.

We do a quick sanity check to determine if the $ID$ is in use in our system or not. If not, then the user's iris scan hash and vaccination record are stored on the blockchain as separate transactions at random intervals of time. We note that the user's iris scan data and personal details are destroyed after use, and are not stored by the CoviChain center or on the blockchain. If on the other hand the $ID$ is found, then we return the user's vaccination record.

When a user wishes to reveal their vaccination status to a third party, they make use of the CoviChain system and provide a scan of their iris which is hashed with LSH algorithm. The result is concatenated with the user's personal details and hashed with the SHA-256 algorithm to recreate the CoviChain $ID$. The $ID$ is then used as an index into the blockchain to locate the user's vaccination record.

A hash function is a one-way function which for any arbitrary input produces a fixed size output (e.g. 256/384/512 bits) \cite{HashFunction}. One of the properties of a hash function is that it is not possible for an adversary to obtain the original input data from the hash output. Another property of hash functions is that even a small change in the input data will result in a large change to the output hash. In particular, hash functions exhibit the ``avalanche effect" \cite{Avalanche}, where a single character change in the input will result in approximately 50\% of the bits flipping in the hash output.

This presents a problem for our system as different iris scanning hardware will produce slightly different scan outputs. This results in a hash that may not match to any hash on the blockchain, or may match to an incorrect hash due to the built-in avalanche effect of hashing algorithms. To counter this, instead of using a traditional hash algorithm we make use of a locality sensitive hashing scheme. In a nutshell, LSH generates ``similar" hashes for similar iris scan inputs. We take each potential match and concatenate them with the user's date of birth and gender, and see if we can match to a CoviChain identifier on the blockchain and retrieve the user's vaccination record.

\subsection{CoviChain Algorithm}

\begin{figure}[ht]
\centerline{\includegraphics[width=18pc]{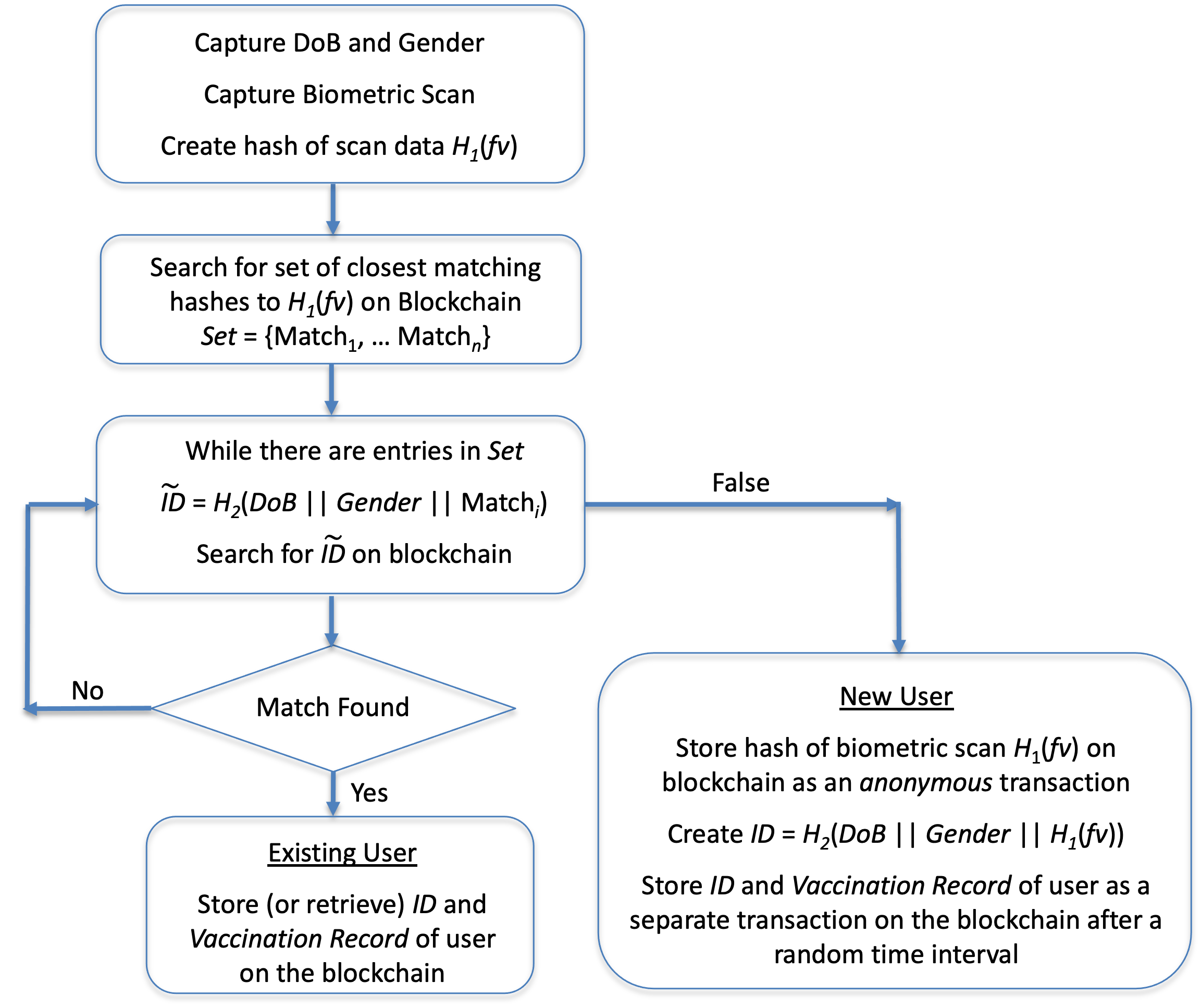}}
\caption{System Overview}
\label{overview}
\end{figure}

Figure \ref{overview} describes the algorithm that we employ in our system. When a user presents themselves to an authorized CoviChain center, they are asked for their \textit{DoB} (dd/mm/yyyy) and \textit{Gender} (male/female/other). In addition, the organization captures a number of scans of the user's iris, and creates a hash ${H_1}(\vect{fv})$ from the feature vector $(\vect{fv})$ extracted from the ``best" biometric scan data. Here $H_{1}$ is a locality sensitive hash function. Our system can combine the user's \textit{DoB} and \textit{Gender} with ${H_1}(\vect{fv})$ to generate a unique 256-bit identifier ($ID$) for the user. In this instance $H_{2}$ is the SHA-256 hash function:

\begin{equation}
ID = {H_2}(DoB~||~Gender~||~{H_1}(\vect{fv}))
\label{eqn1}
\end{equation}

The algorithm tries to match the calculated hash $H_1(\vect{fv})$ with existing ``anonymous" hashes that are stored on the blockchain. It may get back a set of hashes that are somewhat ``close" to the calculated hash, due to fact that we have employed a LSH function. If that is the case, then the algorithm concatenates each returned hash ($Match_i$) with the user's \textit{DoB} and \textit{Gender} to produce $\widetilde{ID}$. It then tries to match each $\widetilde{ID}$ with an $ID$ stored on the blockchain.

If a match is found, then the user is already registered on the system, and has at least one vaccination record. At this point we may just wish to retrieve the user's records or add an additional record, e.g. when a booster dose has been administered to the user. However if we go through the set of returned matches and cannot match any $\widetilde{ID}$ to an existing $ID$ in a vaccination record on the blockchain, i.e. this is the first time the user is presenting to the service, then we store the iris scan hash data $H_1(\vect{fv})$ as an anonymous record on the blockchain, and subsequently the $ID$ and COVID-19 vaccination details for the user as a separate transaction.

In each case the transaction is broadcast at a \textit{random interval} of time on the blockchain peer-to-peer (P2P) network for it to be verified by other authorized nodes in the system, and eventually added to a block on the blockchain. Uploading the two transactions belonging to a user at random intervals of time ensures that the transactions are stored on separate blocks on the blockchain, and an attacker is not easily able to identify the relationship between them.

\subsection{Biometric Authentication}\label{biometric}
When choosing a biometric technique to employ, one has to consider a number of parameters such as: the \textit{security} or strength of the system, the \textit{accuracy} in terms of false acceptance rate (FAR) and false rejection rate (FRR), \textit{permanence} of the biometric which should not change over time, \textit{usability} and \textit{costs} of the system. For the CoviChain system, we chose iris recognition over fingerprint matching, as it provides a higher level of accuracy.

For the purpose of writing this paper we have used the work of Libor Masek \cite{Masek}, which is an open-source implementation of a reasonably reliable iris recognition technique \cite{Kovesi}. Our system takes a scanned image of an iris as the input and outputs a binary feature vector ($\vect{fv}$). Masek’s technique works on grey-scale eye images, which are processed in order to extract the binary template \cite{Chaudhari}. 

\begin{figure}[ht]
\centering
\begin{subfigure}{.45\linewidth}
  \centering
  \frame{\includegraphics[width=.9\linewidth]{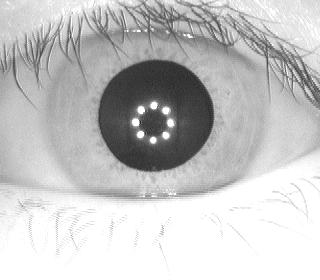}}
  \caption{}
  \label{fig:sub1}
  \vspace{0.5cm}
\end{subfigure}
\begin{subfigure}{.45\linewidth}
  \centering
  \frame{\includegraphics[width=.9\linewidth]{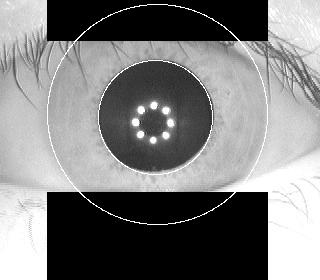}}
  \caption{}
  \label{fig:sub2}
  \vspace{0.5cm}
\end{subfigure}
\begin{subfigure}{0.96\linewidth}
  \centering
  \frame{\includegraphics[width=.9\linewidth]{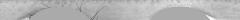}}
  \caption{}
  \label{fig:sub3}
  \vspace{0.5cm}
\end{subfigure}
\begin{subfigure}{0.96\linewidth}
  \centering
  \frame{\includegraphics[width=.9\linewidth]{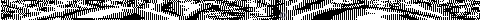}}
\end{subfigure}
\begin{subfigure}{0.96\linewidth}
  \centering
  \frame{\includegraphics[width=.9\linewidth]{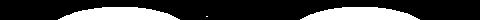}}
  \caption{}
  \label{fig:sub5}
\end{subfigure}
\caption{Masek's Iris Template Extraction Algorithm}
\label{masek}
\end{figure}

Figure \ref{masek} shows the overall process of the iris template extraction technique. First the segmentation algorithm, based on a Hough Transform is used to localise the iris and pupil regions, and also isolate the eyelids, eyelashes and reflections as shown in Figures \ref{masek}a and \ref{masek}b. The segmented iris region is then normalised i.e, unwrapped into a rectangular block of constant polar dimensions as shown in Figure \ref{masek}c. The iris features are extracted from the normalised image by one-dimensional Log-Gabor filters to produce a binary iris template and mask as shown in Figure \ref{masek}d. In summary, the algorithm takes a scanned image as input and outputs a binary feature vector $\vect{fv} \in \{0, 1\}^n$.

\subsection{Locality Sensitive Hashing}\label{lsh}
In order to preserve the privacy of individuals on the blockchain, their biometric data has to be obfuscated before being written to the ledger. Encrypting the data is one possible option. However, hashing is a more efficient mechanism as it is less computationally expensive and requires much less storage. Unfortunately, techniques such as SHA-256 and SHA-3 \cite{SHA} cannot be used for the CoviChain system, since the biometric templates that we extracted above can show differences across various scans for the same individual. Hence using those hash functions would produce completely different hashes. We seek a hash function that generates ``similar” hashes for similar biometric templates. This prompts us to explore LSH, which has exactly this property \cite{Charikar}. 

\begin{figure}[ht]
\centerline{\includegraphics[width=18pc]{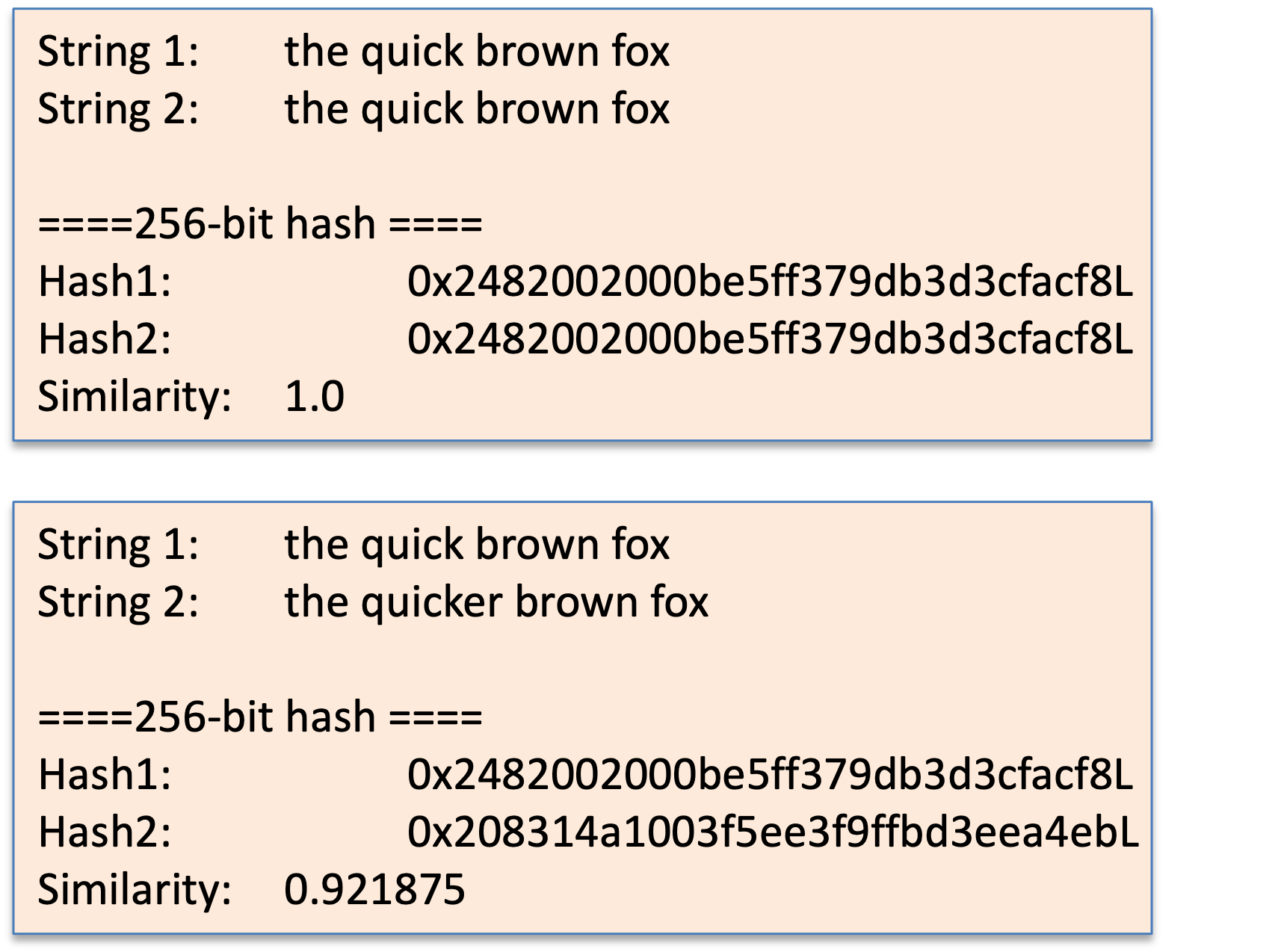}}
\caption{Simhash Output}
\label{simhash}
\end{figure}

Figure \ref{simhash} shows the output from Charikar's Simhash \cite{ASecuritySite} implementation which is a specific implementation of a LSH algorithm. The algorithm takes strings as inputs and the similarity of the resulting hashes can be calculated by finding the percentage of bits that are the same. We can see in the top image that we input two identical strings and that the similarity match is 100\%. Whereas in the second image, we see that the second input string is slightly different from the first, in that we use the work \textit{quicker} as opposed to the word \textit{quick}, with the rest of the words being identical. We can see that since the two strings are similar in most of the bit positions, we get a 92\% similarity match.

In the context of the CoviChain system, we may get a slightly different output from an iris scan procedure when a user is trying to locate their vaccination record on the blockchain. This will result in a different hash output than what was originally stored on the blockchain. Instead of failing to locate an exact match, we look for similar iris scan hashes based on a predetermined threshold. We then take each of those close matches and concatenate the LSH hash with the date of birth and gender details that the user has supplied, and try to match an identifier on the blockchain. If we are successful in matching an $ID$ on the blockchain, we then retrieve the user's vaccination record.

\subsection{Blockchain}
A blockchain is used in the system for immutable storage of individuals’ vaccination records. The blockchain we employ is a permissioned ledger to which blocks can only be added by authorized entities such as hospitals, primary health care centers etc. which operate a full blockchain node. Such entities have to obtain a public-key certificate from a trusted third party and store it on the blockchain as a transaction before they are allowed to add blocks to the ledger. The opportunity to add a new block is controlled in around robin fashion, thereby eliminating the need to perform a computationally intensive proof-of-work (PoW) process.

Any transactions that are broadcast to the P2P network are signed by the entity that created the transaction, and can be verified by all other nodes by downloading the public key of the signer from the ledger itself. CoviChain authorised centres broadcast transactions to the P2P network as they create them. The CoviChain node that is currently designated to create the next block will then collate the transactions that it receives within a certain period of time (e.g. 10 minutes), and adds them to the blockchain in a new block. Any transactions in the mempool that have not been added in the current round are considered for inclusion in the next block.

\begin{figure}[ht]
\centerline{\includegraphics[width=18pc]{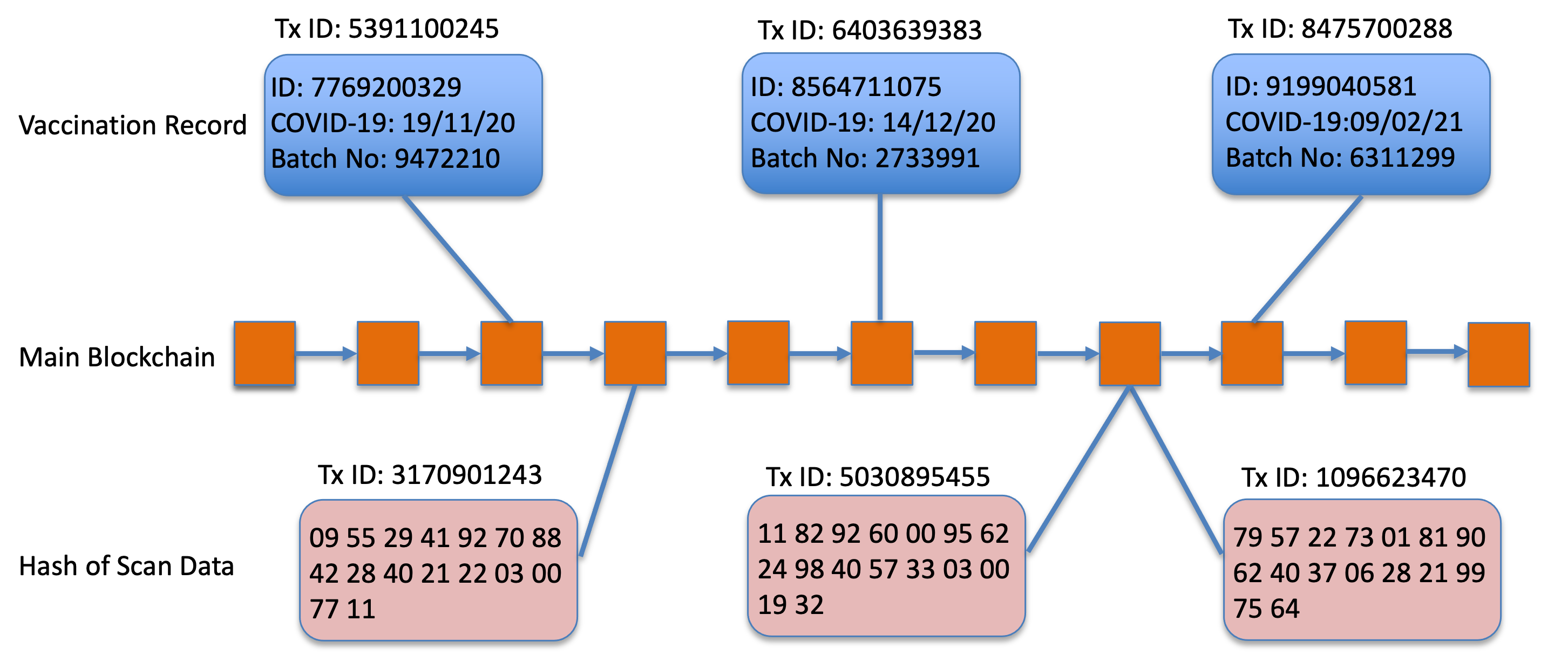}}
\caption{Blockchain Transactions}
\label{blockchain}
\end{figure}

Figure \ref{blockchain} depicts how the CoviChain data is stored on the blockchain, when there are three anonymous transactions (i.e. Hash of Scan Data), and three COVID-19 vaccination record transactions stored on the blockchain pertaining to different users. We make use of a proof-of-authority (PoA) consensus algorithm \cite{Xiao} whereby authorized nodes are able to add transactions to the blockchain. By making use of PoA we no longer require large amounts of computation to be carried out in order to secure the transactions on the blockchain. Note that the storage of the anonymous hash data has to only be carried out once per registered user in the system.

We note that an in-built feature of blockchains is that we can have multiple vaccination records for the same $ID$. For example, when a user receives a booster dose of the vaccine, we can update their vaccination records by adding a new transaction to the blockchain. Since we always access the blockchain from the top block, we are guaranteed to match the most recent transaction belonging to the user first.

\section{Results and Analysis}
The system was evaluated using three different datasets, namely the CASIA-Iris-Interval and CASIA-Iris-Syn datasets \cite{IACAS}, and the IITD iris images \cite{IITD}. Most of the incorrectly segmented irises were removed from the datasets, to prevent them from affecting the matching results. The results for the three different datasets are shown in Table \ref{datasets}.

\begin{table}[ht]
  \centering
  \resizebox{\columnwidth}{!}{%
    \begin{tabular*}{17.15pc}{@{}|p{65pt}|p{20pt}<{\raggedright}|p{33pt}<{\raggedright}|p{38pt}<{\raggedright}|@{}}
    \hline
    \textbf{Dataset} & \textbf{Total Irises} & \textbf{Irises Removed} & \textbf{Irises Remaining}\\
    \hline
    CASIA-Iris-Interval & 2369 & 196 & 2443\\
    \hline
    CASIA-Iris-Syn & 10000 & 2660 & 7340\\
    \hline
    IITD Irises & 2240 & 323 & 1917\\
    \hline
  \end{tabular*}
  }
  \caption{Successfully Segmented Irises}
  \label{datasets}
\end{table}

The accuracy of the system when searching for similar iris scans is considered paramount, as it is the principle step when recreating a user’s unique identifier. Ideally, the system should always be able to locate a user’s record when presented with their iris scan and personal information. However, it is essential that the system never mistakenly identifies a user as someone else. Several metrics are used to evaluate the matching accuracy.

\subsection{Speed of the Iris Extraction Process}
The speed of the system is important for evaluating its viability. Users cannot be expected to wait more than a few seconds to access their vaccination records. If the time required to locate a user’s records is too long, then the system cannot be used in practice.

Results were collected on a laptop with an Intel i7 processor and 16GB of RAM. Table \ref{timings} shows the time taken to run the iris extraction process for each dataset. Based on these results it is apparent that even on a less powerful machine the system should only take a few seconds to extract an iris template from an image.

\begin{table}[ht]
  \centering
  \resizebox{\columnwidth}{!}{%
  \begin{tabular*}{16.55pc}{@{}|p{65pt}|p{20pt}<{\raggedright}|p{32pt}<{\raggedright}|p{32pt}<{\raggedright}|@{}}
    \hline
    \textbf{Dataset} & \textbf{Total Irises} & \textbf{Time Taken (s)} & \textbf{Time per Iris (s)}\\
    \hline
    CASIA-Iris-Interval & 2369 & 3090 & 1.17\\
    \hline
    CASIA-Iris-Syn & 10000 & 10829 & 1.0829\\
    \hline
    IITD Irises & 2240 & 3512 & 1.5679\\
    \hline
  \end{tabular*}
  }
  \caption{Time Taken to Extract Iris Templates}
  \label{timings}
\end{table}

\subsection{Speed and Efficiency of the Blockchain}
\begin{figure}[ht]
\centerline{\includegraphics[width=18pc]{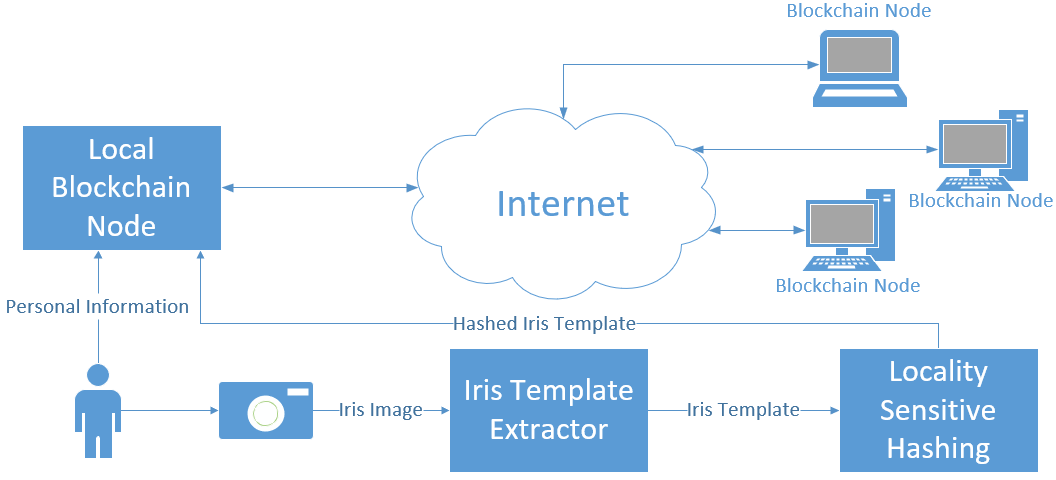}}
\caption{Simulation Setup}
\label{simulation_setup}
\end{figure}

The speed and efficiency with which the system can locate a user’s record was also measured. The evaluation was performed on an Ubuntu Virtual machine running on the aforementioned laptop. It had access to four processors and 8GB of RAM. Figure \ref{simulation_setup} shows our simulation setup in which six nodes were connected to the Ethereum blockchain.

For each user in the dataset one hashed iris scan was stored on the blockchain. Additionally, a unique identifier was constructed for each user using a randomly generated date of birth and gender, along with the stored hashed iris scan. A vaccination record was also created and stored on the blockchain along with the identifier. Iris scans belonging to other users were subsequently inputted into the system to try to retrieve their vaccination records. An acceptance threshold of 0.4 was used for all the datasets when deciding which hashes could have come from the same iris.

\begin{table}[ht]
  \centering
  \resizebox{\columnwidth}{!}{%
  \begin{tabular*}{21.25pc}{@{}|p{65pt}|p{24pt}<{\raggedright}|p{37pt}<{\raggedright}|p{35pt}<{\raggedright}|p{32pt}<{\raggedright}|@{}}
    \hline
    \textbf{Dataset} & \textbf{Records Stored} & \textbf{Searches Performed} & \textbf{Matching Accuracy} & \textbf{Average Time per Iris (s)}\\
    \hline
    CASIA-Iris-Interval & 389 & 2054 & 0.8992 & 0.05047\\
    \hline
    CASIA-Iris-Syn & 947 & 6393 & 0.906 & 0.07519\\
    \hline
    IITD Irises & 407 & 1510 & 0.9795 & 0.01615\\
    \hline
  \end{tabular*}
  }
  \caption{Time Taken to Search Blockchain and Lookup Accuracy}
  \label{search}
\end{table}

The results for the three datasets are shown in Table \ref{search}. The number of matches that were successfully made is reported alongside the average time taken to search the blockchain. Although in some cases the system failed to locate a user’s record, it never misidentified a user as someone else.

\subsection{Blockchain Storage Requirements}
The data usage of the system is also crucial in assessing its viability. The data usage was examined by studying the size of individual blocks and extrapolating the cost. These results assume a block is added every 15 seconds since this is the default time for the Ethereum blockchain. Also, the size of the vaccination records are restricted to 150 bytes. An empty block uses 606 bytes of space. Given there are 86,400 seconds in a day, there will be 5760 blocks added each day. Therefore, the system will use a minimum of 3.49056 MB in one day and 1.27405 GB a year.

The cost to add a million users to the blockchain is now examined. A block with 30 iris hash transactions was found to be of size 5742 bytes. This means each iris hash transaction is approximately 191.4 bytes. Therefore, if one million hashes were stored the estimated space used would be 0.1914 GB. A block with 26 vaccination record transactions was found to be of size 10,102 bytes. This means each record is roughly 388.5385 bytes. Therefore, the estimated space required to store a million records is 0.38854 GB. The total space required to add a million users can thus be estimated at 0.57994 GB.

The maximum space required to add a million users over the course of a year can be calculated to be 1.85399 GB. The actual value would probably be lower since this calculated amount assumes there will be 2,102,400 empty blocks, as well as the blocks required to store the details of the users. In practice, these blocks will overlap which will decrease the overall data usage.

\section{Security Considerations}
Given the complexity of the vaccination passport there are many ways a malicious actor could attempt to attack it. The most important countermeasure is to ensure that no sensitive information is included in the vaccination records. They must be completely anonymous and should only include the bare necessities required to assess someone’s vaccination status. In this way, even if an attacker manages to link a user to their records, they should only be able to deduce their vaccination status. However, this is still considered undesirable. Therefore, several mitigation measures are proposed to minimise the risk of an attacker successfully linking a user to their records.

\subsection{Unique Identifier}
The fact that the records are stored on a publicly accessible blockchain, enables attackers to download them and perform offline attacks. This makes brute force attacks against the system much more viable. Therefore, the CoviChain identifier must be constructed in a manner that renders any potential brute force attack impractical. However, there is a trade-off involved when deciding what personal information should be used to construct identifiers. Using more information increases the entropy, and therefore decreases the likelihood of brute force attacks succeeding and false positives occurring. However, the more personal information that is used, the more details about a user will be leaked if an attack succeeds.

For example, if the personal information solely consists of a user’s date of birth and gender. It would be feasible for an attacker to combine every possible date of birth and gender with every stored hashed iris scan. For every match that is found, the attacker would know the associated user’s date of birth, gender and vaccination records. If the attacker was targeting a specific user and already knew their personal details, it would be easy to locate their records given sufficient time, even without their iris scan.

If some additional personal information was added such as a user’s mother’s maiden name, it would make a potential brute force attack much more computationally expensive. However, attackers could guess particularly common surnames. An alternative countermeasure would be to utilise a secret personal identification number (PIN). This could be in place of or in addition to the personal information. However, the system would then be reliant on users remembering their PINs. This could lead to people losing access to their records. Nevertheless, this may be considered preferable to risking users’ personal information.

\subsection{Iris Anonymity}
It is essential that the system does not leak any information about the users' irises. This will make it much more difficult for an attacker to locate any user's records. Furthermore, the iris itself can be considered sensitive information. Therefore, it is essential that the LSH function be input hiding. This will ensure that information about the users’ irises cannot be leaked from the stored hashes. In \cite{Chaudhari}, the authors show that the S3Hash (a variant of Charikar's Simhash) algorithm is input hiding, provided the entropy of the input vector is sufficiently larger than the size of the hash.

\section{Conclusion}
In this paper we have presented a COVID-19 vaccination passport framework that makes use of biometric iris scan data and personal information, (something you \textit{possess} + something you \textit{know}) to create a unique one-way identifier that can be used to add or locate a person’s vaccination record on a public blockchain, but does not store any personal details about them. Our system balances the needs of governments and the travel and hospitality sectors to allow for the safe reopening of society, while maintaining the privacy of citizens.\\

Conflict of Interest: The authors declare that they have no conflict of interest.


\begin{thebibliography}{1}
\bibliographystyle{IEEEtran}

\bibitem{ACLU} J.Stanley, ``There’s a Lot That Can Go Wrong With ‘Vaccine Passports’", March 2021, \url{https://www.aclu.org/news/privacy-technology/theres-a-lot-that-can-go-wrong-with-vaccine-passports/}.

\bibitem{Liberties} I. Butler and L.Ravo, ``Three Reasons Why a Vaccine Passport for EU Travel Is a Bad Idea", January 2021, \url{https://www.liberties.eu/en/stories/vaccine-passport-free-movement/19061}.

\bibitem{Charikar} S. Charikar, ``Similarity estimation techniques from rounding algorithms", In Proceedings of the 34th Annual ACM Symposium on Theory of Computing, 2002, pp. 380–388.

\bibitem{SHA} Q. Dang, ``Secure Hash Standard", Federal Inf. Process. Stds. (NIST FIPS), National Institute of Standards and Technology, Gaithersburg, MD, 2015, \url{https://doi.org/10.6028/NIST.FIPS.180-4}.

\bibitem{HashFunction} Hash Function - \url{https://en.wikipedia.org/wiki/Cryptographic_hash_function}.

\bibitem{Avalanche} Avalanche Effect - \url{https://en.wikipedia.org/wiki/Avalanche_effect}.

\bibitem{Rui} Z. Rui and Z. Yan, ``A Survey on Biometric Authentication: Toward Secure and Privacy-Preserving Identification," in IEEE Access, vol. 7, pp. 5994-6009, 2019.

\bibitem{Statista} Global biometric technologies market revenue from 2018 to 2027, \url{https://www.statista.com/statistics/1048705/worldwide-biometrics-market-revenue/}.

\bibitem{Masek} L. Masek, ``Recognition of human iris patterns for biometric identification", Final Year Project, The School of Computer  Science and Software Engineering, The University of Western Australia, 2003.

\bibitem{Kovesi} L. Masek and P.Kovesi, ``Matlab source code for a biometric identification system based on iris patterns", The School of Computer Science and Software Engineering, The University of Western Australia, 2003.

\bibitem{Chaudhari} S. Chaudhari, M. Clear, P. Bradish and H. Tewari, ``Framework for a DLT Based COVID-19 Passport", In: Arai K. (eds) Intelligent Computing. Lecture Notes in Networks and Systems, vol 285. Springer, Cham 2021, pp. 108-123.

\bibitem{ASecuritySite} Simhash - \url{https://asecuritysite.com/encryption/simhash}.

\bibitem{Xiao} Y. Xiao, N. Zhang, W. Lou and Y. T. Hou, ``A Survey of Distributed Consensus Protocols for Blockchain Networks," in IEEE Communications Surveys \& Tutorials, vol. 22, no. 2, pp. 1432-1465, 2020.

\bibitem{IACAS} Institute of Automation, Chinese Academy of Sciences, ``Iris image database”, \url{http://biometrics.idealtest.org/}, Accessed: 2021-01-31.

\bibitem{IITD} Indian Institute of Technology Delih, ``Iris image database”, \url{https://www4.comp.polyu.edu.hk/~csajaykr/IITD/Database\_Iris.htm}, Accessed: 2021-03-25.

\end{thebibliography}
\end{document}